\documentclass{article}

\usepackage{arxiv}

\usepackage[utf8]{inputenc} 
\usepackage[T1]{fontenc}    
\usepackage{hyperref}       
\usepackage{url}            
\usepackage{booktabs}       
\usepackage{amsfonts}       
\usepackage{nicefrac}       
\usepackage{microtype}      

\title{Transcending conventional biometry frontiers: Diffusive Dynamics PPG Biometry}

\author{
  Javier de Pedro-Carracedo\thanks{Current address: University of Alcal\'a (UAH), Computer Engineering Department, Alcal\'a de Henares (Madrid), 28871 Spain.} \\
  Departamento de Tecnolog\'ia Fot\'onica y Bioingenier\'ia \\
  Universidad Polit\'ecnica de Madrid (UPM)\\
  Madrid, 28040 Spain \\
  \texttt{javier.depedro@uah.es} \\
   \And
 David Fuentes-Jimenez \\
  Department of Electronics \\
  University of Alcal\'a (UAH) \\
  Alcal\'a de Henares (Madrid), 28871 Spain \\
  \texttt{david.fuentes@depeca.uah.es} \\
  \And
  Ana M. Ugena \\
  Departamento de Matem\'atica aplicada a las Tecnolog\'ias de la Informaci\'on \\
  Universidad Polit\'ecnica de Madrid (UPM) \\
  Madrid, 28040 Spain \\
  \texttt{anamaria.ugena@upm.es} \\
  \And
  Ana P. Gonzalez-Marcos \\
  Departamento de Tecnolog\'ia Fot\'onica y Bioingenier\'ia \\
  Universidad Polit\'ecnica de Madrid (UPM) \\
  Madrid, 28040 Spain \\
  \texttt{anapilar.gonzalez@upm.es} \\
}

\usepackage{amsmath}
\usepackage{amssymb}
\usepackage{multirow}
\usepackage{float}
\usepackage{makecell}
\usepackage{adjustbox}
\usepackage{textcomp}

\usepackage[table,xcdraw]{xcolor}

\usepackage{cite}

\usepackage{cleveref}
\usepackage[caption=false,listofformat=subsimple,labelformat=simple]{subfig}

\Crefname{figure}{Figure}{Figures}

\usepackage{booktabs,dcolumn}
\DeclareMathVersion{nxbold}
\SetSymbolFont{operators}{nxbold}{OT1}{cmr} {b}{n}
\SetSymbolFont{letters}  {nxbold}{OML}{cmm} {b}{it}
\SetSymbolFont{symbols}  {nxbold}{OMS}{cmsy}{b}{n}
\newcolumntype{L}{D{.}{.}{-1}}

\makeatletter
\newcolumntype{B}[3]{>{\boldmath\DC@{#1}{#2}{#3}}c<{\DC@end}}
\newcolumntype{Z}[3]{>{\mathversion{nxbold}\DC@{#1}{#2}{#3}}c<{\DC@end}}
\makeatother

\begin{document}


\maketitle

\begin{abstract}
  In the first half of the 20th century, the first pulse oximeter became available to measure blood flow changes in the peripheral vascular network. The PhotoPlethysmoGraphic (PPG) signal, obtained from a pulse oximeter, is used now to monitor physiological parameters in clinical environments. Over the last decade, it is extending to the biometrics' area. Different methods allow the extraction of each individual's characteristic features from the PPG signal morphology, highly varying with time and physical states of the subject. This paper presents the first PPG dynamic-based biometric authentication system with a Siamese convolutional neural network. Our method extracts the PPG signal's biometric characteristics from its diffusive dynamics, characterized by geometric patterns in the  $(p,q)$-planes specific to the 0--1 test. PPG signal diffusive dynamics are strongly dependent on the vascular bed's biostructure, which is unique to each individual and more stable over time and other psychosomatic conditions than morphology. Besides its robustness, our biometric method is anti-spoofing, given the complex nature of the blood network. Our proposal is trained using a national research study database with 40 real-world PPG signals and compared with eight state-of-the-art methods, achieving the best Equal Error Rate (ERR) and processing times with a single attempt, among all of them.
\end{abstract}


\section{Introduction}\label{sec:introduction}

The relentless outbreak of the pandemic in our lives has put the globalized world in check. Paralysis to which economies across the globe drive reverse, in many cases, by the spread of a latent wave for decades: digitization society. Life will be conditioned by new technologies, an entire online ecosystem whose real impact remains a chimera even among those experts who timidly venture with hasty forecasts \cite{Velikic2020,Dynkin2020,Wang2020}.

The health crisis has perpetrated a double recession, the economic and the social. With intermittent confinements due to the virus's recurrence, the home is our vital refuge, and our interface with society relies entirely on technology. The digital assets of companies focus their efforts on the cloud, factories are automated, and electronic commerce or communication on social networks is growing unstoppably, the virus being one of the most important catalysts of this dynamic.

With the presumable paradigm shift in this vast digital spectrum looming, the global opening towards a fully digital world is imminent. In this time of transformation, structural reforms are peremptory. The role that technology will play in future societies is unquestionable. Virtually everything will connect to everything. However, this profound metamorphosis carries challenges that digital platforms themselves have to face. One of them is to keep the identities of the users of the different services protected, that is, to avoid identity theft so that it can unequivocally verify that a user is who they say they are and not an impostor intruder with clearly fraudulent purposes. Today, the most secure authentication mechanisms are base on biometric methods \cite{Singh2019}. Compared to the traditional use of access passwords, the different biometric identification systems are reliable and free the user from memorizing numerous keys \cite{Sancho2018}. The only access password lies in the user's anatomical characteristic, supposedly exclusive and non-transferable, whose emulation is extremely problematic even for the most seasoned intruders. Face, voice, iris, palm, and finger recognition are already a reality that safeguards socioeconomic transactions. The standard criteria that conventionally met to consider the relevance of a biometric access control system obey \cite{Sanchez-Reillo2000}:

\begin{itemize}
\item Cost of the biometric validation device.
\item Biometric verification speed.
\item Size of the biometric fingerprint that identifies a user.
\item Degree of acceptance by the user.
\item User comfort.
\item Ease of maintenance.
\item Unlinking of police records that may compromise the legal security of the individual.
\end{itemize}

The Big Data growing technology is a direct consequence of the instrumental ease with which one can acquire massive data. In the clinical field, public databases are already available with thousands of physiological samples from healthy individuals and with various pathologies for different age ranges. In this sense, new technological solutions are gradually betting on biological signals that implicitly reveal distinctive anatomical characteristics of each subject, since they record the imprint of the organic systems that intervene in their generation.

The conventional biometric systems focus on the analysis of physical characteristics of an individual, in some cases, highly sensitive to involuntary morphological disturbances---see, for example, a cut on the fingertip undergoes a fingerprint analysis---.  By contrast, biological signals lend themselves to a more robust biometric examination, besides morphological details of the biological signal waveform, dynamic peculiarities by the expected functional response of the physiological system of interest evaluate.

In the last decades, a preliminary diagnostic examination of an individual's health status has been entrusted on many occasions to the pertinent clinical analysis and monitoring, using non-invasive methods, of biological signals generated by the human body \cite{Moraes2018}. Among the different biological signals that are usually measured today, one, particularly, deserves special consideration, the PPG signal. The PPG signal reproduces blood volume changes that occur in the peripheral microcirculation \cite{Sancho2018}. The electro-optical technique that allows measuring these volumetric variations calls photoplethysmography (PPG) \cite{Elgendi2012a}.

Since Alrick Hertzman, an American physiologist, devised the first photoelectric photoplethysmograph in 1937 \cite{Hertzman1937}, although rudimentary, recent technological advances provide devices, as modern pulse oximeters, increasingly smaller, lighter, and with a marked tendency to market themselves as wireless devices at a very affordable price \cite{Castaneda2018,Peart2018}. An essential aspect of the PPG technique lies in its low sensitivity to the sensors' location, which gives versatility to photoplethysmography for its application in many areas, such as health, sports, or the agri-food industry. Appearance due to the electronic simplicity, the cost-benefit ratio, the ease of signal acquisition, and, mainly, its non-invasive character \cite{Allen2007,Elgendi2012,Sviridova2015}. Unlike other biological signals that require bulky measurement equipment, or even accessories, such as gels (EEG) or electrodes (ECG), the PPG signal requires quite modest electronics. Electronics and optoelectronics uncomplicated, encourage the construction of small pulse oximeters, easily integrable into smart devices \cite{Yadav2018}. A pulse oximeter consists of a light emitter and a photodetector. The photodetector senses changes in light absorption resulting from arterial blood pulses (pulse signal or PPG) when a light beam passes through or reflects in human tissue \cite{Webster1997}.

The PPG signal is widely used in clinical settings to monitor physiological parameters related to the cardiorespiratory system \cite{Dhar2018}. The PPG signal is complex. It composes an AC component---peripheral pulse synchronizes to each heartbeat---; and a quasi-DC part that varies slowly, due to respiration, vasomotor activity, and vasoconstrictor waves \cite{Meredith2011}. The mutual coupling between the different components is intricate and operates at different timescales to regulate blood volume based on physiological needs.

In this work, we use the PPG signal dynamics as a biometric reference of any individual. In this sense, we focus our attention on the geometric distribution of the PPG signal's diffusive behavior, according to the $(p,q)$-plane proposed by the 0--1 test \cite{Gottwald2004,Gottwald2005,Bernardini2015}. We believe that the PPG signal's diffusive dynamics are unique to each individual since the diffusion constant of blood flow is subject to the structural configuration with which each individual has been endowed \cite{Pedro-Carracedo2019}. A whole complex network of arterioles and capillaries transports blood from the heart to the rest of the body thanks to the heart's driving force and synchronized with the respiratory rhythm. Although variations in the diffusive dynamics of the PPG signal can indeed hide point or progressive pathological abnormalities, such as, for example, physiological deterioration as a result of ageing, specific congenital characteristics remain practically unchanged. In this regard, the blood dynamics (hemodynamics) in the peripheral capillary branching stand as ideal candidates for identifying an individual's biometric pattern.

The database used comes from 40 students between 18 and 30 years old, non-regular consumers of psychotropic substances, alcohol, or tobacco. The students were selected to participate in a national research study to evaluate how the stress reflects on biological signals \cite{Aguilo2015, Arza2018}. Signals were captured from the middle finger of the left hand and sampled at a frequency of 250 Hz \cite{Aguilo2015}, so sampling time is $\Delta t=4$ ms with the psychophysiological telemetric system ``Rehacor-T'' version ``Mini'' from Medicom MTD Ltd \cite{Aguilo2015}. In this experimental setting, the PPG signals capture the blood microcirculation and are a true reflection of the peripheral capillary branching that each individual presents.

Each subject's credentials, their identity, is collected in blood flow dynamics through the peripheral capillary network. Its falsification is very difficult because of capillary network's intricacy and the complexity which involves blood flow driven by the cardiorespiratory system. Furthermore, significant detail is that any biometric system based verifying PPG signal's diffusive dynamics requires the individual's vital integrity. Someone, not without a negligible effort, could imitate the particular capillary morphology of an artificial finger. Still, it would be practically impossible to reproduce the diffusive dynamics that blood flow undergoes when circulating through that capillary structure, given the contribution of many subsystems that do nonlinearly make up the cardiovascular system.

Our proposal involves the recognition of dynamic patterns that characterize the diffusive behavior of the PPG signal. As an authentication mechanism \cite{Sancho2018}, the biometric architecture consists of two stages: In the first phase, \textit{the enrollment phase}, 12 s of PPG signal are acquired from each individual using a pulse oximeter. These signal fractions are preprocessing to obtain several $(p,q)$-planes representative of each subject, the biometric pattern. From these $(p,q)$-planes, the neural network extracts 51,200 characteristics that encapsulate each individual's biometric pattern and conveniently stored in memory. Afterward, in \textit{the verification phase}, 12 s PPG signal acquire from anyone who wishes to verify their identity, proceeding to their preprocessing. Through a classifier and their comparison with the rest of the registered biometric patterns, authenticate the user's identity that requests it. The use of 12 s of PPG signal in each of the phases of the system is because it is the time necessary to obtain three consecutive segments of PPG signal (4 s or 1000 points each one), with their respective $(p,q)$-planes from the user, to be recorded or verified. Additionally, the use of 12 s to verify a user's identity enables applying this system in real environments, since, with this not too long time, achieves accuracy above 90\%.

The rest of this paper is organized as follows. \Cref{sec:stateofart} condenses the advances made in the application of the PPG signal as a biometric marker over the past few years. \Cref{sec:Summaryofcontributions} outlines the relevance of this paper within the context of PPG-based biometry. \Cref{sec:01test} describes the theoretical foundation, in the mathematical framework of the 0--1 test, that underpins the biometric potential of the geometric patterns traced by the diffusive behavior of the PPG signal. In \cref{sec:CNNarch} our novel proposal for a biometric classifier based on convolutional neural networks is explained in detail. \Cref{sec:RESULTS} shows the obtained results, both graphically and numerically, for various experimental settings. Also, in this section, we analyze and interpret the obtained results. Finally, in \cref{sec:RESULTS}, we shortly outline the conclusions drawn from this study, which serve as the basis for future work.

\section{State of the art}\label{sec:stateofart} 

With the development of biometrics during the 20th century, according to its definition in \cite{Paulsen2019}: ``Measurable physical characteristics or personal behavioral traits used to identify or verify the identity of an individual'', began by conforming to the old paradigm of facial recognition and of fingerprints. Nevertheless, continued progress in the area of image processing and analysis has fostered the exploration of more sophisticated biometric system designs \cite{bio5,Yin2019} (for a sound review of classical biometric approaches and their evolution over time, you are referred to \cite{Jain2004}).

So far, in the 21st century, the development of biometric pattern recognition systems have evolved enormously, broadening its application spectrum in the context of morphological analysis, as reflected between the proposal of the anatomical characterization of the hand geometry in \cite{Sanchez-Reillo2000} and the made by \cite{bio2} concerning 3D palmprint modelling. The same is true for other biostructure patterns as disparate as geometric characterization of ear \cite{bio4}, of iris \cite{Wang2019}, of the eye as a multimodal biometric system \cite{bio1}, and of the distribution of veins in a finger \cite{bio3} or on the wrist \cite{Garcia-Martin2020}.

However, in the 21st-century particular attention must be paid to the use of biological signals like biometric markers, in addition to morphological and behavioral characteristics. In this regard, worth highlighting biometrics studies involving the analysis of electrocardiographic (ECG) and encephalographic (EEG) signals \cite{Singh2012}, to which could be added biometric applications that obtain the biological signals from the galvanic response of skin (GSR), electromyogram (EMG), electrooculography (EOG), and mechanomyogram (MMG), among others \cite{Pal2015}.

Over the years, technological advances have simplified the acquisition of biological data; somehow, \textit{traditional biometric systems} (TBS) have been increasingly giving way to \textit{wearable biometric systems} (WBS), and, thus, to new methodological approaches to computing and validating biometric patterns \cite{Blasco2016}. Accordingly, new biometric technologies are gradually abandoning the rigidity imposed by a stationary and static analysis of biometric patterns \cite{Kim2018} towards biometric patterns adapted to the variations that the biological signals may undergo over time---the so-called \textit{adaptive biometric systems} \cite{Pisani2019}---. In the particular case of the PPG signal, biometric patterns are strongly conditioned to physiological alterations, such as physical activity, emotional states, and time intervals in which measurements do. An apart from the impact of the different noise sources coupled in the PPG signal acquisition procedure \cite{Yadav2018}, mainly when the PPG signal is obtained from a camera or of wrist-worn PPG collected in an ambulant environment \cite{Biswas2019}.

Focussing now on the matter at hand, the first documented reference to the PPG-based biometric system dates back to Gu \textit{et al.}'s research work in 2003 \cite{Gu2003}. In all the works that use the PPG signal as a biometric reference, specific biomarkers correspond to features implicitly or explicitly extracted from the signal waveform. For example, time-domain features acquired from PPG signal's first and second derivatives for biometric identification \cite{Kavsaoglu2014}, or approximating each PPG signal as a sum of Gaussians, and using the parameters in a discriminant analysis framework to distinguish individuals \cite{Sarkar2016}, or also defining the waveform of the PPG signal in five consecutive PPG cycles \cite{Spachos2011} or from 22 cycles \cite{Lee2015} parametrically. One of the latest work is related to the non-fiducial and fiducial approaches for feature extraction with supervised and unsupervised machine learning classification techniques \cite{Karimian2017}. Another on the simultaneous PPG signal acquisition using different wavelengths, the video camera detectors allow extracting the color segment (e.g., red, green, and blue) \cite{Patil2018}. In all PPG-based biometric models, a negative aspect is the nature non-stationary of the PPG signal over time, which prevents the stable identification of an individual's biometric patterns.

\section{Summary of Contributions}\label{sec:Summaryofcontributions}
 
We present four contributions to the state-of-the-art in PPG-based biometry within the DNN framework.
 
  First, as a general concept of the system, we propose a biometric system based on the diffusive dynamics of the PPG signal with a DNN design adapted to diffusive images and a specific biometrics method. Our proposal technically bases on the preprocessing of the 0-1 test \cite{Nicol2001} and the Siamese residual network structures.
  
Second, our proposal only needs 12 s of the PPG signal to verify the identity or register users in terms of its inputs. The proposal does not need to retrain for each user because it dynamically adapts to its template pattern, being this the one the registered by the user, making the registering and identifying task easier than in the other state-of-the-art approaches.
   
Third, in this work, we moved away from the morphological analysis of the PPG signal to extract an individual's biomarkers, and we now concentrate attention on its diffusive dynamic behavior, geometrically mapping on a 2D plane---do not be confused with extracting the PPG signal from an image \cite{Sun2016} or remote sensing---, according to the theoretical approach introduced by the 0--1 test \cite{Pedro-Carracedo2019}, as advanced in \cref{sec:introduction}.

Finally, we present quantitative experimental results that our proposal outperforms state-of-the-art accuracy, robustness, and computation time. These results only include real-world, non-synthetic PPG signals. Our proposal arises from the study carried out in the national research project \cite{Aguilo2015,Arza2018} wherein different biological signals of 40 individuals are analyzed, in one basal session and the other with induced stress, using standard convolutional neural networks \cite{Ruiz-Orozco2019}. In light of preliminary findings, biometric patterns obtained through our PPG-based biometric system show promising stability to changes in the individual's emotional state and the time lag, about one month, between samples of the PPG signal collected, unlike it can see in \cite{Sarkar2016}.

 \section{0--1 test}\label{sec:01test}

In the analysis of dynamical systems, one of the key aspects is to characterize the dynamic behavior present in the physical system's response under study. As such, the dynamics of the response do not provide direct relevant information on the internal physical structure from which the response derives. Still, it does provide at least its operational complexity, which is crucial in evaluating its correct functioning and its more or less adaptability to unforeseen situations in the context of physiological systems.

In an experimental setting, observables usually get from the physical system under consideration so that the observables make measurements at regular time intervals. An observable is any physical quantity that can be measured. The measurements or observations organize in what is known as time sequences, and then each observable gives rise to a scalar time sequence. These temporal successions are submitted to the pertinent scrutiny to unravel the system's functionality that generated them. We could define a state vector in phase space if we measured all the observables that contribute to given dynamical system evolution. In the field of physiological systems, it is widespread to work with univariate temporal sequences or scalar temporal sequences, in which only the measurements of an observable are available. With a single observable, it is possible to obtain information on the state of the system, since each of them usually contains information from the others, given the mutual coupling between them, whether linear or non-linear.

The initial motivation for the 0--1 test was to have a method to apply directly to a scalar temporal sequence to identify the presence of chaotic dynamics without having to resort to other more complicated techniques that require a deep level of knowledge for its correct application and interpretation \cite{Gottwald2004,Gottwald2005,Gottwald2009}. Given its easy implementation, its increasing popularity has sparked the interest of countless scientific disciplines in an excessive race to detect chaos anywhere \cite{Bernardini2015}. However, beyond the initial scope of the 0--1 test and its many applications, we believe that one of the phases of the test is surprisingly useful in the field of biometrics; specifically, the auxiliary trajectory of the two-dimensional Euclidean group (\textit{the Fourier transform series}), or $p$-$q$ diagram or $(p,q)$-plane \cite{Nicol2001}, which underlies the dynamics of the physical system.

The 0--1 test cornerstone construction of an extended dynamic serves a two-dimensional Euclidean group $\mathbf{SE}(2)$ \cite{Nicol2001}. The elements of $\mathbf {SE}(2)$ form rigid displacements, that is, a translation and a rotation, in some two-dimensional affine Euclidean plane---the $(p,q)$-plane---that, in principle, it does not relate in topological terms to the state space in which the dynamics of the system unfold. However, parameters that characterize rigid transformations depend at all times on the current state of the system and, therefore, a certain equivalence relationship manifest between the dynamics of the physical system under study and the dynamic evolution that the trajectory described by the elements of $\mathbf {SE}(2)$ in the $(p, q)$-plane.

\begin{figure}[tbhp]
\centering
\includegraphics[width=0.5\linewidth]{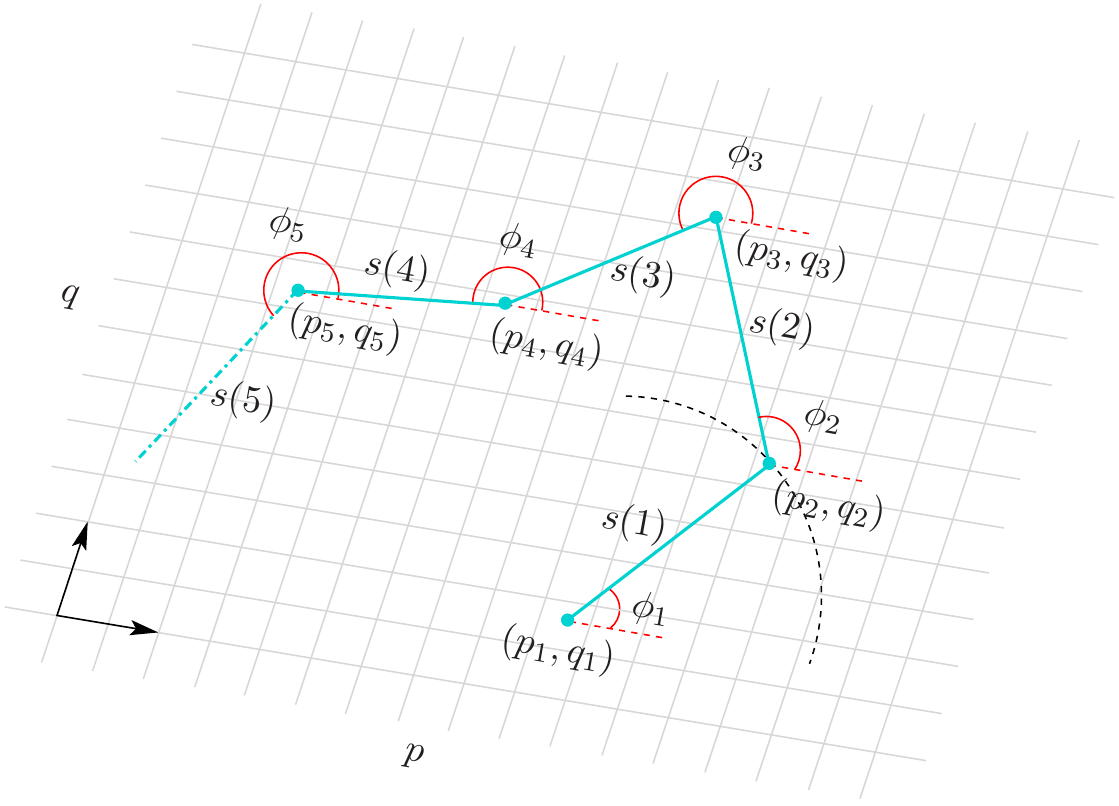}
\caption{Descriptive construction of the auxiliary trajectory in the $(p,q)$-plane.}\label{fig:FSE1}
\end{figure}

The 0--1 test requires as input a scalar time sequence of $N$ observations $s(n)$, for $n=1,2,\dotsc,N$, where $s(n)$ is a one-dimensional observable of the underlying dynamical system. According to the rigid transformations' parameterization, the extension of the dynamics characterizes by $s(n)$ forces to define three scalar quantities $(p,q,\phi)$. An element or point on the $(p,q)$-plane characterize by its position on the plane, whose coordinates are $(p,q)$, although its evolution, a change in coordinates, is driven (\textit{forcing term}) by the dynamic evolution of $s(n)$ according to

\begin{align}\label{eqn:SE01}
p_{n+1} &= p_{n}+s(n)\cos \phi_{n},\nonumber \\
q_{n+1} &= q_{n}+s(n)\sin \phi_{n},  \\
\phi_{n+1} &= c+\alpha s(n),\nonumber
\end{align}
where parameters $c,\alpha\in \mathbb{R}$.

The evolution of any point on the $(p,q)$-plane describes a trajectory called the auxiliary trajectory since it reproduces an indirect or complementary evolution of the true dynamics observed in the system. The auxiliary trajectory involves an angular rotation $\phi_ {n}$ with respect to a circumference of radius $s(n)$ centered on the point $(p_ {n}, q_ {n})$, as shown in \Cref{fig:FSE1}.

Somehow the auxiliary trajectory derives from a diffusive process in which the diffusion dynamics are forced or driven by the $s_ {n}$ observations. In the presence of noise, for dynamic simplicity, $\alpha$ usually assigns a value of 0 \cite{Gottwald2005, Gottwald2009} so that \Cref{eqn:SE01} reformulates as

\begin{align}\label{eqn:SE02}
p_{n} &= \sum_{k=1}^{n} s(k)\cos(kc),\nonumber \\
q_{n} &= \sum_{k=1}^{n} s(k)\sin(kc),  \\
\phi_{n} &= cn,\nonumber
\end{align}
where the angle of rotation $\phi_ {n}$ increases at a uniform rate governed by the value of $c$. Furthermore, since the parameter $c$ participates in the trigonometric function's argument, it is pertinent that $c\in [0,2\pi)$.

Although the theory underlying the dynamic extensions is based on the dynamics' asymptotic behavior, an interesting consequence of this focuses attention on the limited nature of auxiliary trajectories in the $(p,q)$-plane. That is, how the auxiliary trajectory evolves spatially in the $(p,q)$-plane if the trajectory is circumscribed in an area delimited from the $(p,q)$- plane or inexorably diffuses by the $(p, q)$-plane in the same way that a Brownian motion unfolds \cite{Gottwald2009}. The 0--1 test quantifies by the computation of an indicator whether the auxiliary trajectory is bounded, in which case it reflects the presence of regular dynamics, or not sublinearly bounded, in which case it manifests the presence of chaotic dynamics. This inductive argument is the base of the 0--1 test; a more in-depth description goes beyond this article's objective. Readers are referred to this method's original work, widely referred to in the scientific literature in the last decade \cite{Gottwald2004,Gottwald2005,Gottwald2009,Gottwald2009a}.

This article uses young and healthy individuals' PPG signals, more precisely, the geometric configurations described by their auxiliary trajectories in the $(p, q)$-plane to identify individuals biometrically by considered convolutional neural networks. As already stated in another article \cite{Pedro-Carracedo2019}, the dynamic richness of the auxiliary trajectories of the PPG signals, for a range of values of the parameter $c$ that prevents the appearance of spurious phenomena, reveals the inherent functional complexity to signal dynamics, to which multiple conveniently coupled physiological subsystems contribute. The coordinated action of these subsystems is responsible for homeostatic regulation of the cardiorespiratory system at all times. However, despite the certain global similarity that the auxiliary trajectories of PPG signals may have at first glance, closer scrutiny of each individual shows distinctive signs that could hide more or less diagnostic severe pathologies, and, more invariably, the inalienable character of the anatomical and functional configuration of the cardiorespiratory system of each subject. Future work in this line could yield promising results. The approach presented here uses the dynamic analysis of diffusive processes provided by the 0--1 test, the diffusive geometry represented in the $(p, q)$-plane, as an additional preprocessing layer. This layer allows the PPG signals acquired in the time domain to convert into geometric maps according to the $(p, q)$-planes of the input PPG signals. The data used to train and evaluate the neural network address the PPG signals of 40 young and healthy individuals from a national study \cite{Aguilo2015, Arza2018}. The neural network uses each individual's geometric maps to establish the similarity or divergence between different subjects and could validate or deny an input user's identity to the identities stored and authorized by the authentication system. On the other hand, this article also examines the proposed method's reliability and performance following current biometric standards. As far as we know, diffusive dynamics, the cornerstone of the 0--1 test, of a biological signal have never used to extract biometric characteristics, which gives this work a new operational perspective in physiological biometrics.

\section{Classifier}\label{sec:CNNarch}
In this article, we propose an approach based on convolutional neural networks to identify users through their PPG signals, so that it is possible to know if two different PPG signals correspond to the same user. The system proposed here receives two-time segments of PPG signals, 1,000 points on each, as inputs, of which the first will be the standard segment and the second the segment of the user to compare. The system delivers as output a matching score normalized to the interval $[0,1]$, which defines the degree of agreement between the two incoming PPG segments, namely, if the two input segments belong to
the same user, matching score closer to 1, or not, matching score closer to 0. As far as the authors know, no other study considers the design of a biometric system using the diffusive dynamic information of a biological signal, according to the $(p,q)$-planes provided by a preliminary phase of the 0--1 test. In binary images, its output patterns feed a convolutional neural network that subsequently identifies the incoming users concerning those already registered. On the other hand, other studies try to solve the biometrics problem from other points of view in the PPG signal \cite{Kavsaoglu2014,Sarkar2016,Lee2015} and through other types of biological patterns \cite{bio1,bio2,bio3,bio4,bio5,Kumar2013,Liang2020}.

\begin{figure}[tbhp]
	\centering
	\includegraphics[width=0.95\linewidth]{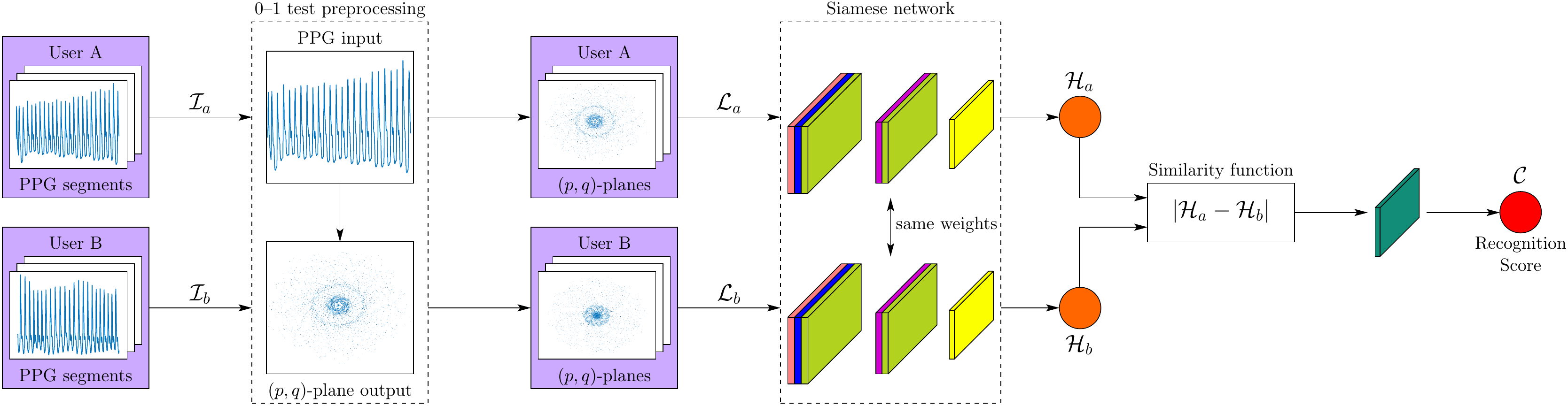}
	\caption{System's architecture schematic overview.}\label{fig:diagCNN}
\end{figure}

\subsection{Architecture}
This article proposes a non-conventional network, as we can see in \Cref{fig:diagCNN}, whose main architecture base on a Siamese network whose main trunk is characterized by a fully-connected encoder, a multiscale architecture with residual connections based on the guidelines of Szegedy \textit{et al.} \cite{inceptionResnet}. Fully connected encoder architectures are those traditionally used in classification tasks such as \cite{Simonyan2015,Szegedy2015}. In the Siamese configuration case, this is well-known for its use in one-shot learning and image verification \cite{Koch2015SiameseNN}. In addition to these layers and architectures somewhat better known in the field, adding a layer of its own to the system performs preprocessing based on the diffusive behavior peculiar to the PPG signal dynamics \cite{Pedro-Carracedo2019} highlighted as a new contribution to this article.

The reason for using Inception-ResNet-V1 \cite{inceptionResnet} architecture as the main branch of the Siamese network is due to its well-known capacity as a classifier and its characteristics concerning its previous versions and competing networks:

\begin{itemize}
\item Reduction of architectural bottlenecks for \cite{inception1,inception2}, because the neural network works better if the dimensional input changes are not too drastic. Large dimensional changes can cause a significant loss of information called a ``representational bottleneck''.
\item Use of factoring methods to reduce the computational complexity of the convolutions used.
\item Use of residual connections between the inputs and outputs of the blocks used. These connections prevent the loss of information and improve the stability of the gradients when training.
\item Use of batch normalization to immunize the network to some extent against changes of scale, reduce training time, and avoid covariance displacement.
\end{itemize}

\setcellgapes{5pt}\makegapedcells
\begin{table}[tbhp]
\caption{Propose CNN detailed architecture. All the sub-blocks that belongs to the original Inception-ResNet architecture can be found in the seminal paper \cite{inceptionResnet}.}\label{tab:TCNNarch}
	\centering
	\begin{adjustbox}{max width=\linewidth}
		\begin{tabular}{|c|c|c|c|}
			\hline
			\textbf{Layer number} & \textbf{Type}  & \textbf{Output size} & \textbf{Configuration} \\ 
			\hline
			$1_{\text{A}}$ & Input & $(1000,3)$ & --- \\ 
			\hline
			$1_{\text{B}}$ & Input & $(1000,3)$ & --- \\ 
			\hline
			$2$ & 0--1 test preprocessing & $2\cdot(299,299,3)$ & Siamese \\ 
			\hline
			$3$ & Stem  & $2\cdot(35,35,256)$  & Siamese \\
			\hline
			$4$ & $5\times$ Inception-ResNet-A & $2\cdot(35,35,256)$ &  Siamese  \\
			\hline
			$5$ & Reduction-A & $2\cdot(17,17,896)$ & Siamese \\	
			\hline
			$6$ & $10\times$ Inception-Resnet-B & $2\cdot(17,17,896)$ & Siamese \\
			\hline
			$7$ & Reduction-B & $2\cdot(8,8,1792)$ & Siamese \\		
			\hline
			$8$ & $5\times$ Inception-Resnet-C & $2\cdot(8,8,1792)$ & Siamese \\	
			\hline			
			$9$ & Similarity function & $(8,8,1792)$ & --- \\	
			\hline
			$11$ & Flatten & $114688$ & --- \\	
			\hline
			$12$ & Dense & $1$ & --- \\	
			\hline
			$13$ & Sigmoidal activation & $ 1$ & --- \\	
			\hline
		\end{tabular}
	\end{adjustbox}
\end{table}

The basic structure of the proposed system takes the form of a network combining 1D information (PPG signals) and 2D information ($(p,q)$-planes of PPG segments). This structure contains two distinct phases. The first phase consists of a preprocessing layer based on the characteristic $(p,q)$-planes of the 0--1 test. This phase will have as input six segments of the PPG signal from two users, three belonging to a registered user $\mathcal{P}_{r1},\mathcal{P}_{r2},\mathcal{P}_{r3}$, and the rest to a candidate user $\mathcal{P}_{c1},\mathcal{P}_{c2},\mathcal{P}_{c3}$, not necessarily different. Once these signal segments enter the 0--1 test preprocessing layer, their signals are featured with this process and six output matrices are obtained $\mathcal{I}_{r1},\mathcal{I}_{r2},\mathcal{I}_{r3}$, and $\mathcal{I}_{c1},\mathcal{I}_{c2},\mathcal{I}_{c3}$, which can be represented as an image $\mathcal{I}=[\mathcal{I}_{1},\mathcal{I}_{2},\mathcal{I}_{3}]$, which represent the patterns corresponding to the PPG signals of those users.

The second phase will use as input these six output matrices obtained in the previous phase, in two matrices with three channels each, since each user has three matrices assigned to him. This phase consists of a Siamese network whose architecture base on \cite{inceptionResnet}. This network will use a single coding branch to process the two input matrices separately, with the same trunk and sharing the same weights. Some coded output features $\mathcal{F}_{r}$ and $\mathcal{F}_{c}$ will be obtained for each of the input matrices. Once features obtain, a relation function of these characteristics uses to quantify the error between them and quantify how similar these users are to each other. This error function represents the $L^{1}$-norm between the vectors of characteristics previously obtained. Once the $L^{1}$-norm standard obtains between the characteristics vectors, these will go through a final fully connected binary classification layer. A sigmoidal activation used to obtain a final $\mathcal{C}$ score between 0 and 1, which quantifies how similar or different the evaluated users are. The architecture can observe in detail in \Cref{tab:TCNNarch}.

\subsection{Training}
In this section, we will proceed to explain the training. The first thing that needs to be explained is the training data used in this manuscript. We use for training 40 real-world PPG signals from different individuals obtained from a national research study \cite{Aguilo2015,Arza2018}. Each of the signals is divided into 150 randomly chosen segments of the signal, resulting in 6000 different PPG segments, and because we use three images per user, resulting in $\frac{6000}{3}^{\frac{6000}{3}}$ possible training combinations.

In this step, we train the neural network with data from real-world biological signals, divided into training, validation, and test sets, composed of 60\%, 20\%, and 20\%, respectively, of the data from the database. These division ranges commonly chosen to ensure that almost half of the data used for evaluation. The training parameters explain below.

\subsubsection{Optimizer}
The optimizer used is \textit{Adam} or Adaptive Moment Estimation \cite{Kingma2014}. This optimizer is an excellent alternative to the conventional Stochastic Gradient Descent (SGD). It combines the advantages of two previous alternatives \cite{Duchi2011,Dauphin2015}, creating a new approach that uses the averages of the first and second moments of the gradient to dynamically adapt the learning rate. When we talk about the learning rate, we refer to parameters that define how much and how fast the system learns in each period. This parameter is crucial and can produce great learning problems if it does not choose correctly. A very high learning rate can produce divergence in training, while a very low rate can easily fall into local training minima or take a long time to complete. When we talk about \textit{Adam}'s adaptive capabilities, we mean that in practical terms, it starts with a user-defined learning rate, but after that, it modifies the learning rate through unsupervised training, this allows the use of an adaptive training ratio, but that depends strongly on the batch size and how noisy the input is. The training ratio initially used is $10^{-4}$. In addition to \textit{Adam}'s functionality, a callback called early stopping has been employed in this training. This tool allows the best weight settings to save that the system has achieved throughout the training. In order to achieve this, the tool uses the metrics and losses obtained by the model in each period, so that it can save only the best model of all the periods trained---the network trained during 100 epochs with a batch size of 5. However, a predefined number of epochs used, as we have commented before, the early stopping will keep the best of them. The total training time, on a GPU NVIDIA GeForce GTX 1080, has been of 9 hours.

\subsubsection{Loss function}
The proposed convolutional neural network uses as input two PPG signal segments $\mathcal{I}_{a}$ and $\mathcal{I}_{b}$, while as output, it uses a binary classification vector $\mathcal{C}$. This binary classification task's proposed loss function is the cross-entropy (CE), as indicated in \Cref{eqn:E01CE}, which evaluates the differences between \textit{Ground Truth} and predictions to provide an output score associated with the input signals' similarity. In classical machine learning, this loss function has been widely used to solve the problems associated with a binary classification between distributions, being $p(x)$ the correct distribution and $q(x)$ the estimated one, in such a way that it allows to associate a similarity score for those distributions.

\begin{equation}\label{eqn:E01CE}
\text{CE}(p, q)=-\sum_{\forall x} p(x) \log (q(x)).
\end{equation}

Binary cross-entropy measures the classifier's capacity under study, whose output is a classification level that associates the input to the distribution of interest. The more this classification level decreases, the more the cross-entropy losses increase. The perfect classifier would have zero cross-entropy with a maximum classification level. Usually, this loss function is used in neural networks accompanied by an output activation according to it. In binary cross-entropy, activation is sigmoid, which places the output score level in the interval $[0,1]$, with a smooth transition.

Finally, we will explain step by step how the training of the neural network takes place. The problem to be solved by this system is a binary classification problem with only two possible classes, class 0, which indicates that the input PPG segments of branch A and branch B do not belong to the same user, and class 1, which indicates that these segments belong to the same user. Each of the predefined training segments generated with a specific output label links these input segments A and B to an output classification, allowing the system to learn how to differentiate or associate the input segments of different users. Once in the training process, a random batch generator will use allowing choose 3 PPG signal segments belonging to user A from among the 40 PPG signals used and another 3 PPG signal segments belonging to user B, once again randomized, so that if these two users coincide an output label will be applying with class 1, while if not, it will be associated with class 0. This generator allows guaranteeing the highest possible variability, greatly enriching the training and providing it with generality. Once the batches generate, \textit{Adam} optimizer is used to train the system in the task of recognizing similar users. It is necessary to emphasize that the division of data between training, validation, and testing has been 60\% for training data and 40\% for validation and test data. After training, this system will be evaluated on the previously defined test sets, configured in different ways, as explained in the following sections. Each of these operating modes may require retraining of the system so that a correct ablation study can guarantee the proper functioning and conditions of this operation in different operative situations or work regimes.

\section{Results}\label{sec:RESULTS}

In this section, we show the biometric potential of the diffusive dynamics of the PPG signal. To do this, we explore its operational feasibility under different experimental conditions to mimic its effectiveness in possible real-life scenarios. 

\subsection{Preprocessing}\label{ssec:PPGfiltering}

In practice, the PPG signal is usually impaired by many common noise sources during the signal acquisition process, such as motion artifacts, sensor movements, breathing, etc, as well as for the discretization error (truncation error) involved in normalizing the input signal amplitudes. A common and direct mechanism to mitigate the effect of noise is to submit the PPG signal to a bandpass filter. For filtered PPG signals, we use a Butterworth bandpass filter tuned to different cutoff frequencies. Anything below 0.5 Hz can be attributed to baseline wandering, while anything above 8 Hz is high-frequency noise \cite{Slapnicar2019}, though some studies have reported clinical information up to 15 Hz \cite{Allen2007,Reisner2008}. To examine the impact that this early preprocessing has on the learning and the final performances of our biometric system, we study the following variations:

\begin{enumerate}
\item Raw data: in this first mode, the PPG signals are not pre-processed and transferred directly, as they were acquired, to the 0--1 test preprocessing layer (see \Cref{fig:diagCNN}), where once segmented, they convert to diffusive geometric maps.
\item Filtered data [0.1--8 Hz]: in this second mode, the PPG signals, before moving to the 0--1 test preprocessing layer, are filtered with a Butterworth bandpass filter with cutoff frequencies at 0.1 and 8 Hz, and the amplitudes are not normalized.
\item Filtered data [0.5--8 Hz]: in this third mode, the PPG signals, before moving to the 0--1 test preprocessing layer, are filtered with a Butterworth bandpass filter with cutoff frequencies at 0.5 and 8 Hz, and the amplitudes are not normalized.
\item Filtered data [0.5--8 Hz] and normalized: in the latter mode, the PPG signals, before moving to the 0--1 test preprocessing layer, are filtered with a Butterworth bandpass filter with cutoff frequencies at 0.5 and 8 Hz, and the amplitudes normalized to the $[0,1]$ interval.
\end{enumerate}

\subsection{Metrics}
Once the modalities in which the experimentation will carry out fixed, the metrics used to evaluate the proposed system's performance are explained:

\begin{itemize}
\item \textbf{Precision-Recall curve}. The precision-recall curve depicts the precision vs. the sensitivity (recall) for different operating points (matching score or threshold values). The closer the curve is to the upper right corner (the area under the curve is closer to 1), the more precise and sensitive the system behaves. The accuracy evaluates how often the output is correct (positive). An accurate system is very finicky validating a legitimate user, i.e. in an accurate system it is unlikely that an intrusive user will be admitted as valid, but it is also possible that legitimate users will be rejected (false negatives). Sensitivity assesses how permissive the system is, i.e., in a highly sensitive system, it is very unlikely that a valid user will be rejected, but it is also possible that unregistered users will be admitted as valid (false positives).
\item \textbf{ROC (Receiver Operating Characteristic) curve}. The ROC curve depicts sensitivity vs. FPR (false positive rate). The closer the curve is to the upper left corner (the area under the curve is closer to 1), the more sensitive the system behaves without increasing FPR. In short, the ROC curve graphically represents TPR (true positive rate) vs. FPR (false positive rate) for different operating points (matching score or threshold values).
\item \textbf{F$_1$ score-Threshold curve}. The F$_1$ score-Threshold curve complements the information provided by the Precision-recall curve. F$_1$ score is a joint and overall metric that brings together the Precision and Recall values in a unique metric (precision and recall harmonic mean) that allows us to estimate the stability of the system's performance for different threshold values. In a stable and high-performance system, the range of threshold values for which the curve remains almost constant and close to 1 is virtually a flat line over the whole range.
\item \textbf{Equal Error Rate (EER)}. The equal error rate or crossover error rate (CER) is a metric concerning biometric authentication systems that determines a working threshold where FPR (false positive rate) and FNR (false negative rate) are the same. The point where these decision errors cross define the working point, and the lower the crossover rate, the higher the system's accuracy. At the experimental level, EER is used as a metric to compare different biometric authentication techniques.
\end{itemize}

Usually, a high decision threshold identifies an accurate model with a very low FPR (false positive rate); a low threshold value indicates a high sensitivity (too permissive, with a very low FNR (false negative rate)). The precision-recall and ROC curves help us to find the equilibrium threshold. In our case, the criteria for selecting the optimal threshold comes from the EER, but the F1 score-Threshold curve tells us if variations of the optimal threshold upwards or downwards would dramatically affect the system performance.  Based on the results we will see later, the precision-recall and ROC curves' equilibrium threshold would not be so critical, as the system's stability has a wide operating margin for a not insignificant range of working thresholds.

\subsection{Experimental conditions}\label{ssec:ALLEXP}

In this section, we present two different modalities of experiments. In the first modality, we use 60\% of users to train and the 40\% remaining to test. This approximation allows us to show the system's generalization capacity, with better applicability to real systems, showing its results in new user patterns isolated from the trained users. The second modality and the most used in the published biometry papers \cite{Sancho2018,Patil2018,Yadav2018,Karimian2017,Sarkar2016,Lee2015,Kavsaoglu2014,Spachos2011} use 60\% of all data from all users to train and 40\% to test, and this means that the used patterns are isolated but belongs to the same users, which leads to a certain extent to the presence of similarities.

\subsubsection{Leaving 40\% of users out of training}\label{sssec:EXP1}

In this first experiment, 60\% of users are taken as the training set and a test set with the other 40\% users. This way, the network is trained with 24 users and tested with 16 users never seen before. This experiment allows us to completely isolate 16 users so that the network has never seen a similar pattern in the training phase. Therefore, the register of authorized users does not record the biometrics ID of the 16 users who keep out. 

\begin{figure}[thbp]
	\centering
	\includegraphics[width=0.5\linewidth]{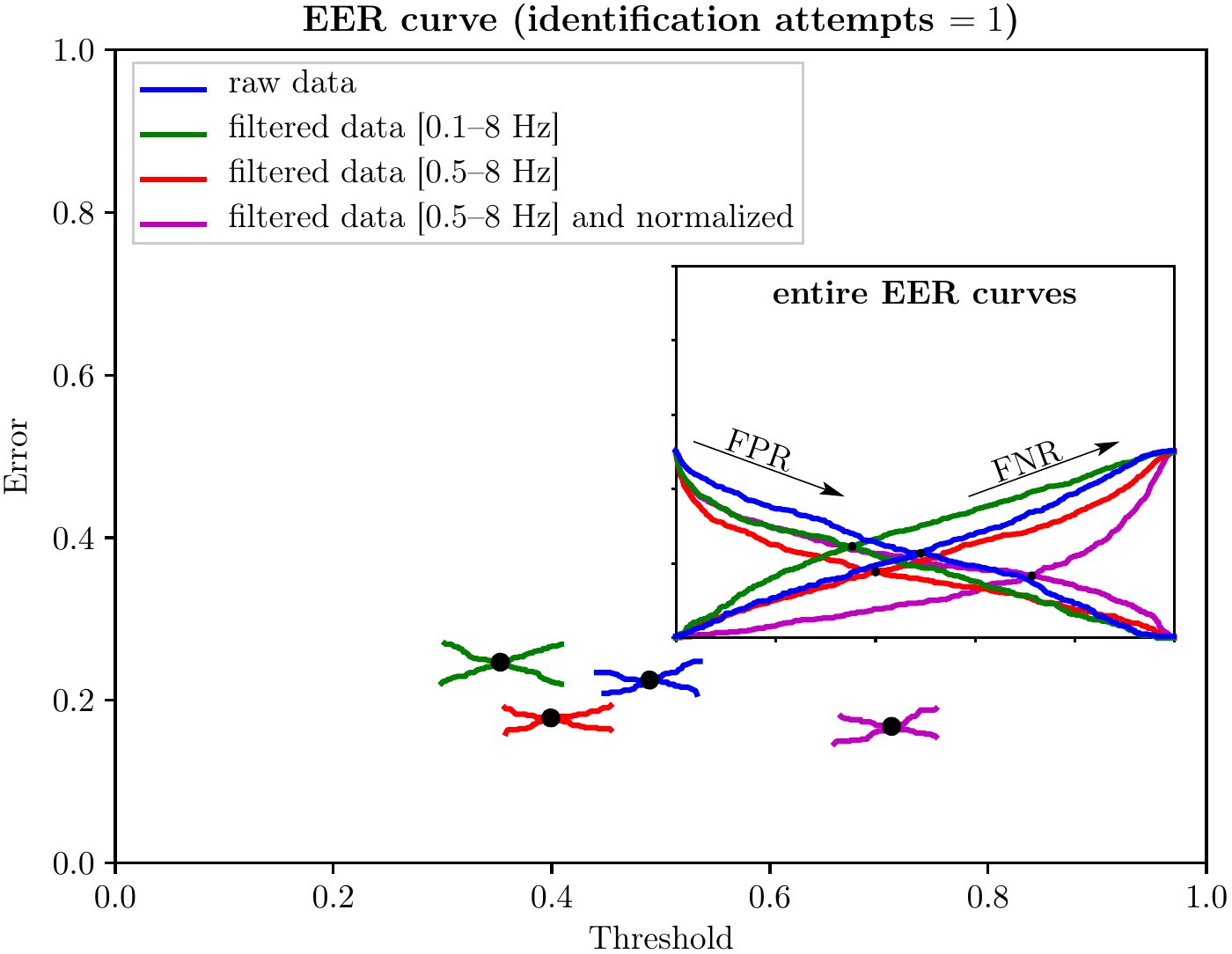}
	\caption{Minimum Equal Error Rate (EER) for different input PPG signal preprocessing modalities. The inset shows the entire EER curves as well as FPR (false positive rate) and FNR (false negative rate) trends for different threshold values.}
	\label{fig:EER_EXP1}
\end{figure}

\Cref{fig:EER_EXP1} shows the different EERs for all the input PPG signal modalities used (cf. \S\ \ref{ssec:PPGfiltering}). For raw data and filtered data in the range of 0.1 to 8 Hz, the network's discriminating power is penalized by the noise present in the signal, which distorts and blurs the diffusive geometrical patterns in the $(p,q)$-planes. As filtering narrows its bandpass in the range of 0.5 to 8 Hz, the impact of noise attenuated, and the diffusive geometric pattern becomes clearer, allowing the network to discriminate between different users' biometric patterns more easily. If, besides, PPG signal normalized to $[0,1]$ interval, once filtered in the range of 0.5 to 8 Hz, a slight reduction in EER is seen. This effect is because the signal's normalization improves the numerical quantification, and the diffusive geometric patterns trace a better structural resolution, making it easier to extract the biometric features.

\begin{figure}[tbhp]
\centering
\subfloat[]{\includegraphics[width=1.96in]{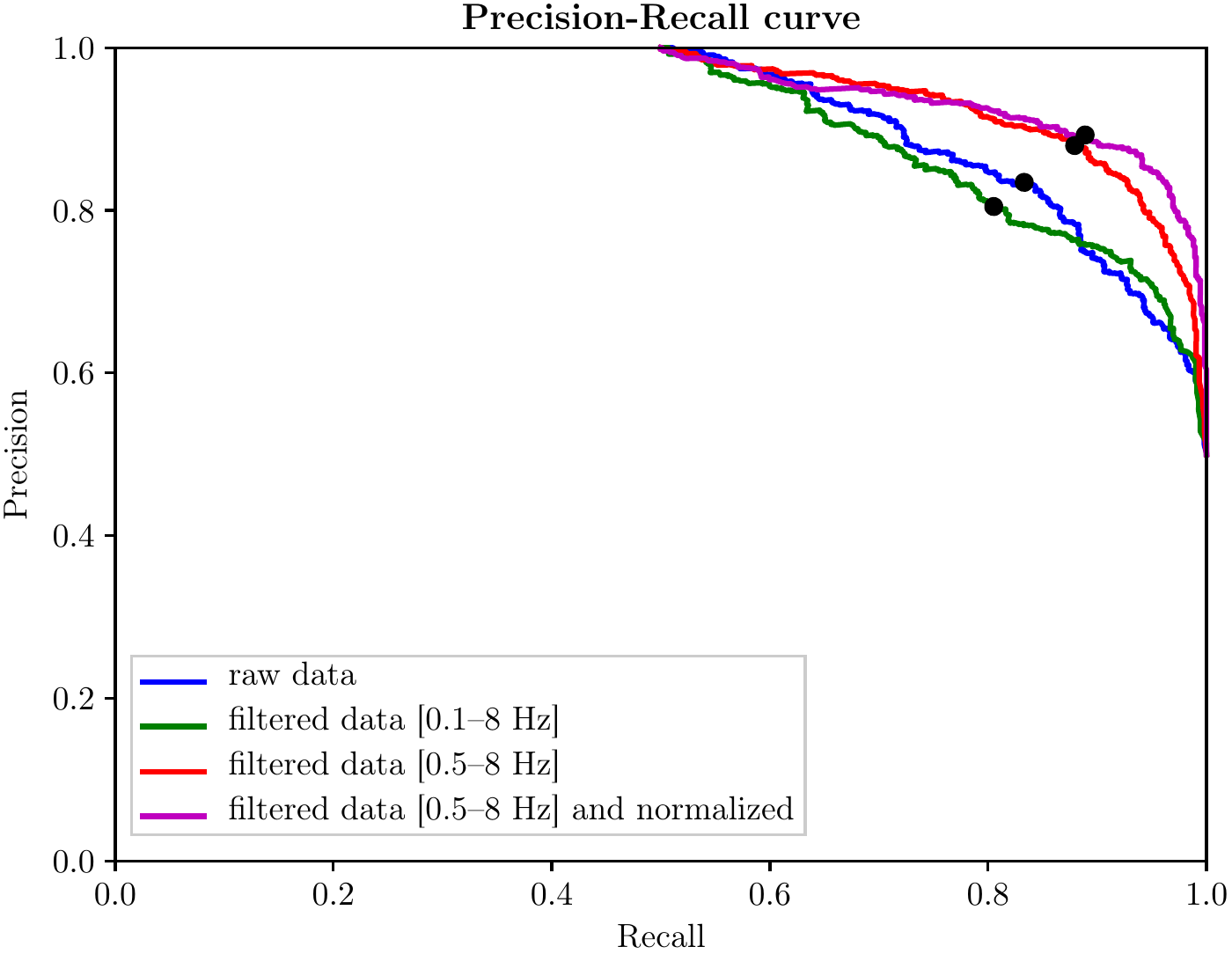}
\label{fig:PR_EXP1}}
\subfloat[]{\includegraphics[width=1.96in]{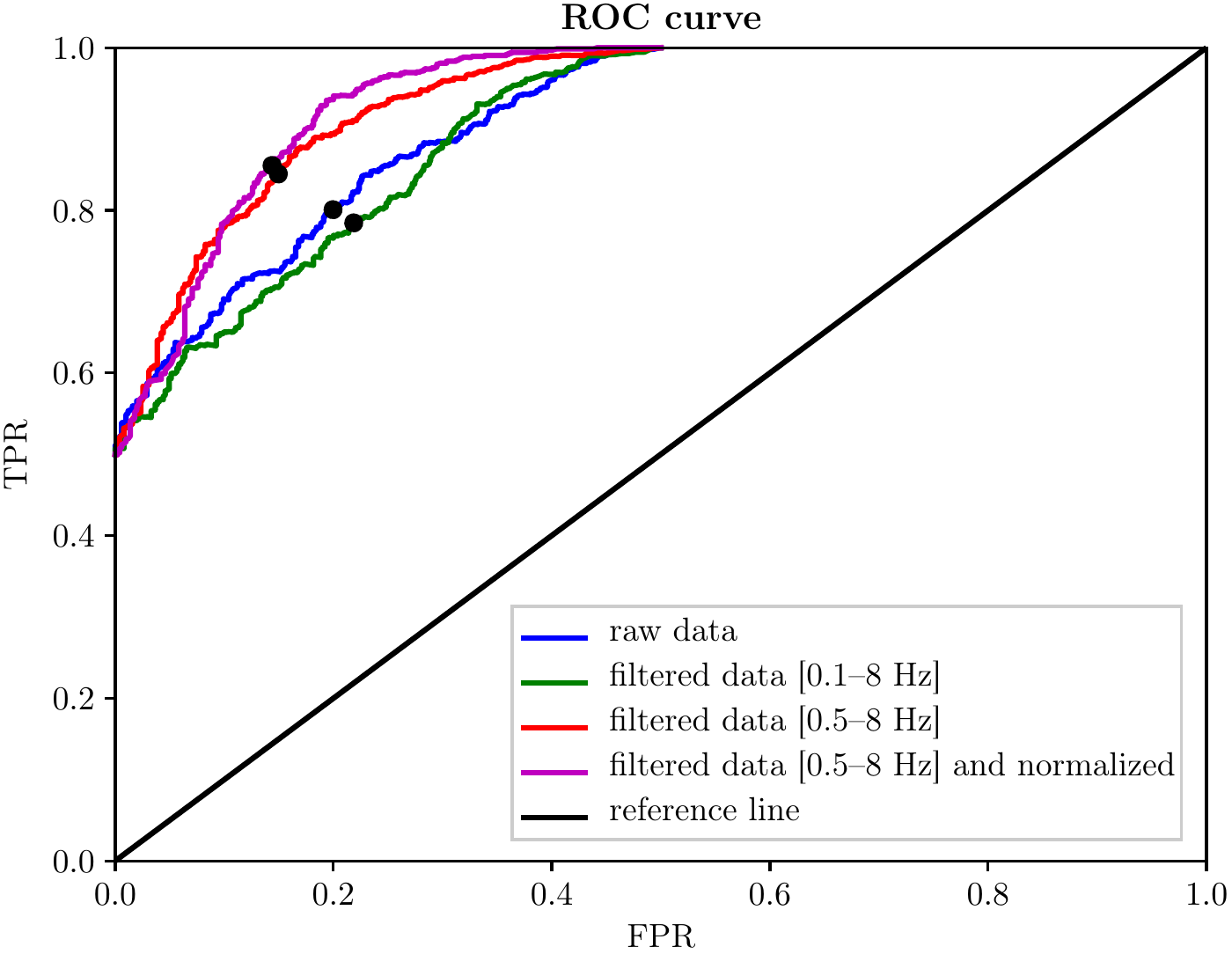}
\label{fig:ROC_EXP1}}
\subfloat[]{\includegraphics[width=1.96in]{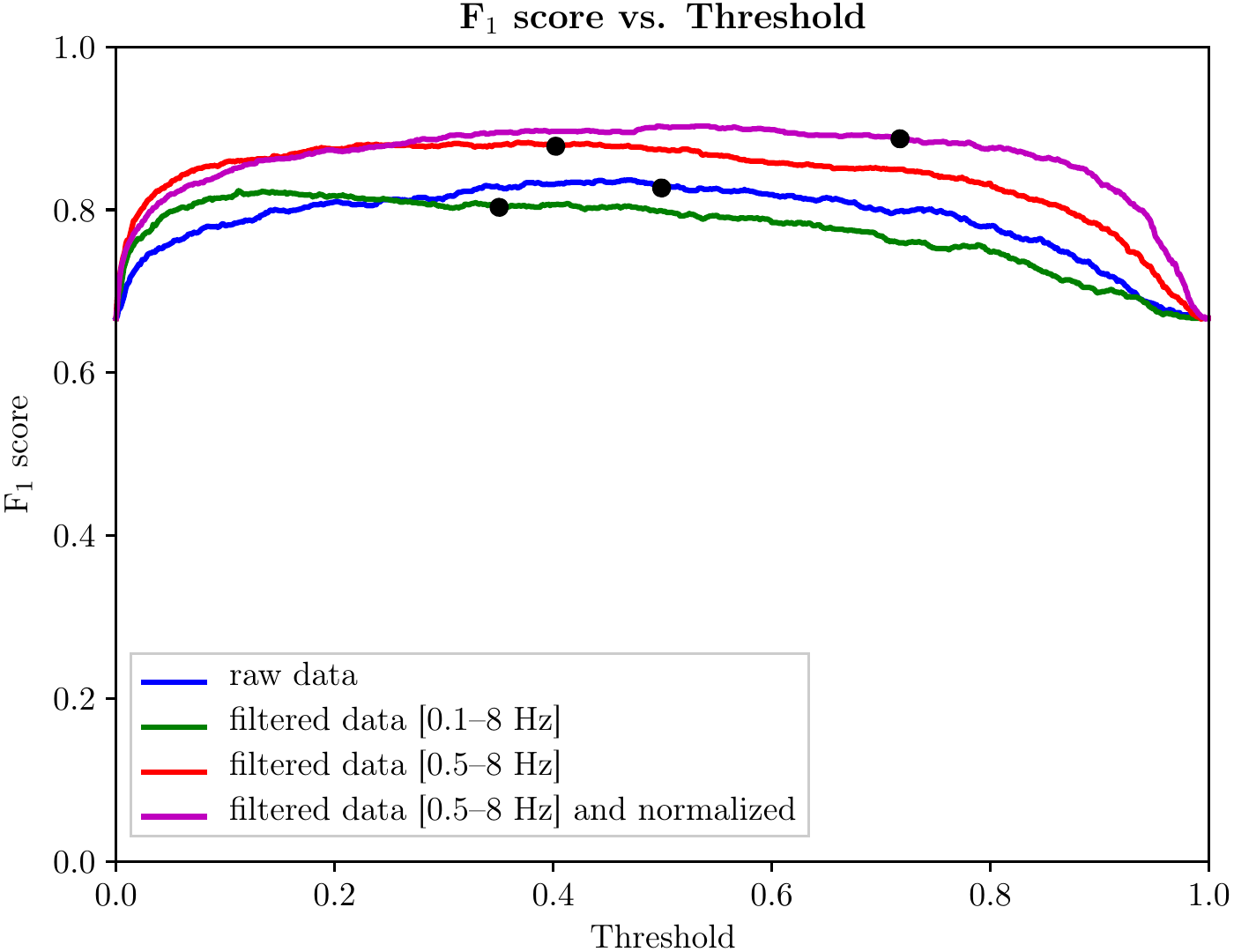}
\label{fig:F1_EXP1}}
\caption{Functional efficiency curves in case leaving 40\% of users out of training. The working points of the EER curve (see \Cref{fig:EER_EXP1}) are tagged with the symbol {\large\textcolor{black}{\textbullet}}: (a) Precision-Recall curve; (b) ROC curve; (c) F$_1$ score-Threshold curve.}
\label{fig:FCEXP1}
\end{figure}

From the EER curve can be measured the working points for each of the preprocessing modes. These working points can use to obtain other performance measures, as shown in \Crefrange{fig:PR_EXP1}{fig:F1_EXP1}. It can be seen that for raw data and filtered data in the range of 0.1 to 8 Hz, the functional efficiency curves, that is, Precision-Recall, ROC, and F$_1$ score-Threshold curves, behave quite similarly. However, as illustrated in \Crefrange{fig:PR_EXP1}{fig:F1_EXP1}, the filtering in the range of 0.5 to 8 Hz provides a significant enhancement in system operating performance, especially about the stability of the working point, pointed out by the F$_1$ score-Threshold curve, much higher than the raw data and filtered data in the range of 0.1 to 8 Hz. Unlike in terms of EER curve in functional efficiency curves, the benefit of $[0,1]$ interval normalization, once filtering the data in the range of 0.5 to 8 Hz, is remarkable. On the one hand, there is a marked improvement in performance for high thresholds, and, on the other hand, in the F$_1$ score-Threshold curve, the working point is much more stable than in any of the other modes.

\setcellgapes{5pt}\makegapedcells
\begin{table}[ht!]
\caption{Performance metrics for all the input PPG signals modalities use in case leaving 40\% of users out of training. The thresholds refer to the optimal classification thresholds where EER is minimal for each modality (preprocessing) considered.}\label{tab:TEXP1}
\centering
\begin{adjustbox}{max width=\linewidth}
\begin{tabular}{|L|L|L|L|L|}
\hline
\multicolumn{5}{|c|}{\textsc{raw data}} \\
\hline
\multicolumn{1}{|c|}{\textbf{Precision}} &
\multicolumn{1}{|c|}{\textbf{Recall}} &
\multicolumn{1}{c|}{\textbf{F$_1$ score}} & \multicolumn{1}{c|}{\textbf{Threshold}} & \multicolumn{1}{c|}{\textbf{Equal Error Rate (EER)}} \\	
\hline
0.82 & 0.82 & 0.82 & 0.48 & 0.22 \\
\hline
\multicolumn{5}{|c|}{\textsc{filtered data [0.1--8 Hz]}} \\
\hline
\multicolumn{1}{|c|}{\textbf{Precision}} &
\multicolumn{1}{|c|}{\textbf{Recall}} &
\multicolumn{1}{c|}{\textbf{F$_1$ score}} & \multicolumn{1}{c|}{\textbf{Threshold}} & \multicolumn{1}{c|}{\textbf{Equal Error Rate (EER)}} \\	
\hline
0.80 & 0.80 & 0.80 & 0.37 & 0.23 \\
\hline
\multicolumn{5}{|c|}{\textsc{filtered data [0.5--8 Hz]}} \\
\hline
\multicolumn{1}{|c|}{\textbf{Precision}} &
\multicolumn{1}{|c|}{\textbf{Recall}} &
\multicolumn{1}{c|}{\textbf{F$_1$ score}} & \multicolumn{1}{c|}{\textbf{Threshold}} & \multicolumn{1}{c|}{\textbf{Equal Error Rate (EER)}} \\	
\hline
0.89 & 0.89 & 0.89 & 0.40 & 0.19 \\
\hline
\multicolumn{5}{|c|}{\textsc{filtered data [0.5--8 Hz] and normalized in $[0,1]$ interval}} \\
\hline
\multicolumn{1}{|c|}{\textbf{Precision}} &
\multicolumn{1}{|c|}{\textbf{Recall}} &
\multicolumn{1}{c|}{\textbf{F$_1$ score}} & \multicolumn{1}{c|}{\textbf{Threshold}} & \multicolumn{1}{c|}{\textbf{Equal Error Rate (EER)}} \\	
\hline
0.90 & 0.90 & 0.90 & 0.73 & 0.18 \\
\hline
\end{tabular}
\end{adjustbox}
\end{table}

\Cref{tab:TEXP1} shows the performance metrics of the experiment whereby 40\% of users are left out of training. 

\subsubsection{Leaving 40\% of data out of training}\label{sssec:EXP2}

In this second experiment, the results are shown taking 60\% of the total data as the training set, which therefore includes all users, covers all users, and a test set with the remaining 40\% of the data. Meaning that the network handles $(p,q)$-planes for all users in the training phase, but in a different way than they will be treated for testing, even though they are undoubtedly related to the specific users' biometric patterns.

This experimental framework sets out a particular setting in which the registered users' database is already known, and therefore, no new user enrollments will occur. All users are well-known to the network since they have enrolled previously.

\begin{figure}[tbhp]
	\centering
	\includegraphics[width=0.5\linewidth]{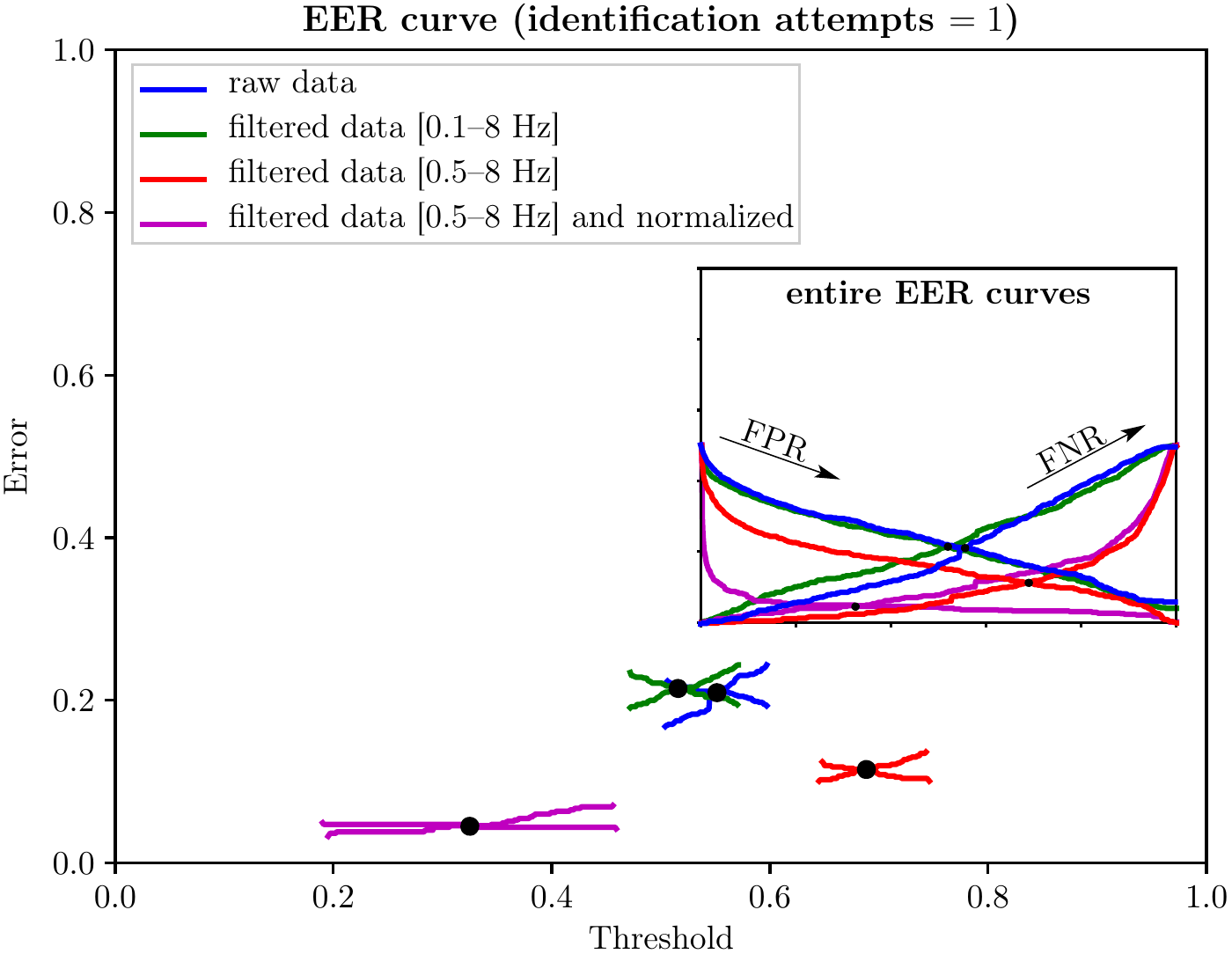}
	\caption{Minimum Equal Error Rate (EER) for different input PPG signal preprocessing modalities. The inset shows the entire EER curves as well as FPR (false positive rate) and FNR (false negative rate) trends for different threshold values.}
	\label{fig:EER_EXP2}
\end{figure}

\Cref{fig:EER_EXP2} shows the different EERs for all the input PPG signals modalities used (cf. \S\ \ref{ssec:PPGfiltering}). For raw data and filtered data in the range of 0.1 to 8 Hz, the network's discriminating power is similar to that obtained in the preceding experimental framework (cf. \S\ \ref{sssec:EXP1}, \Cref{fig:EER_EXP1}). The noise present in the signal, which distorts and blurs the diffusive geometrical patterns in the $(p,q)$-planes, is a critical constraint on the biometrics system's operational capability. 

Nevertheless, contrary to what appears in \Cref{fig:EER_EXP1}, for filtering in the range of 0.5 to 8 Hz, the network offers high efficiency, with a significant reduction of EER. If, besides, to PPG signal normalization in the interval [0, 1], it applies a filter in the range of 0.5 to 8 Hz, reach the lowest EER, very close to zero (6\%, as indicated in \Cref{tab:TEXP2}). With such credentials, it is already clear how a properly preprocessing of the input PPG signals can positively influence the biometric system's final throughput.

\begin{figure}[tbhp]
\centering
\subfloat[]{\includegraphics[width=1.96in]{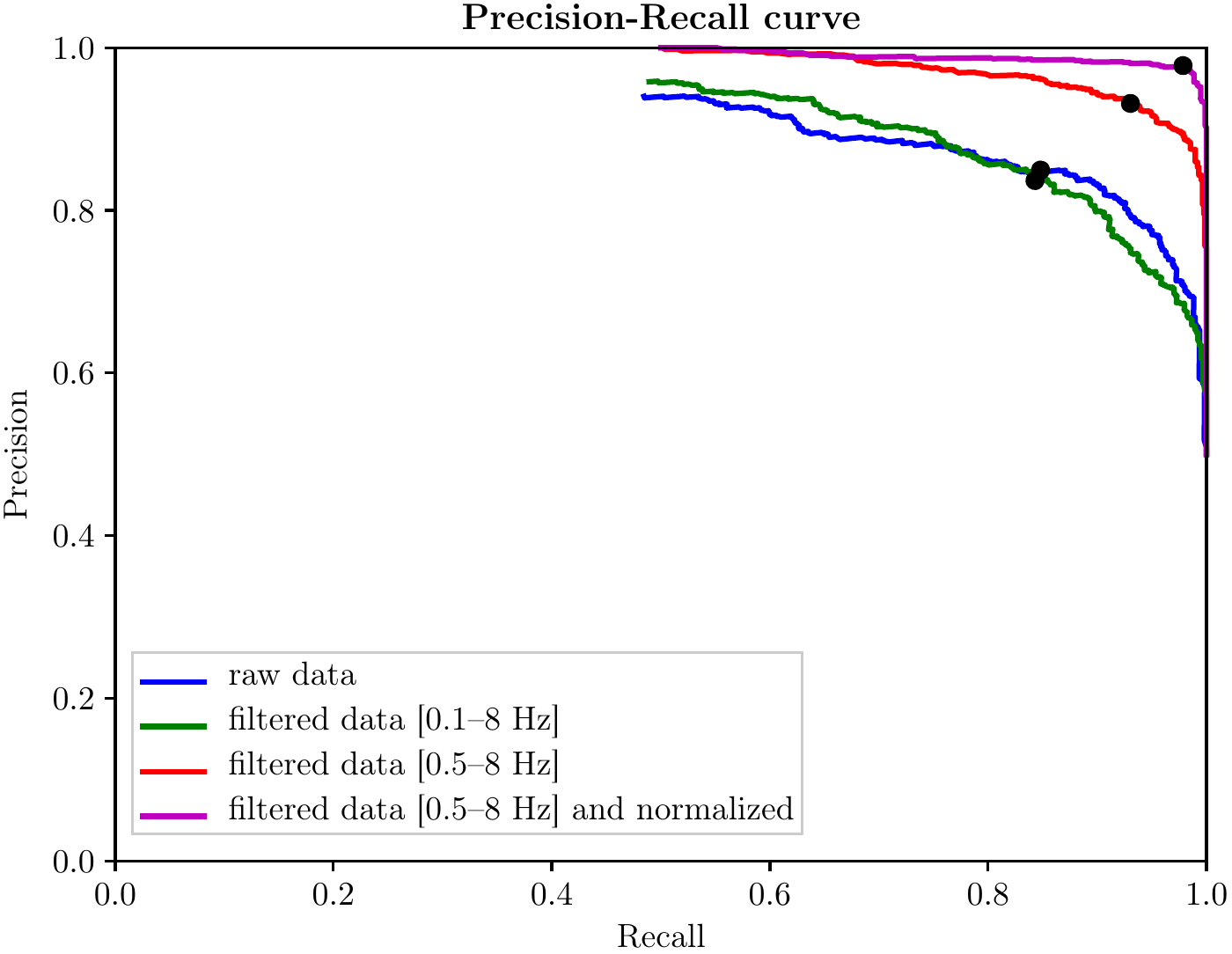}
\label{fig:PR_EXP2}}
\subfloat[]{\includegraphics[width=1.96in]{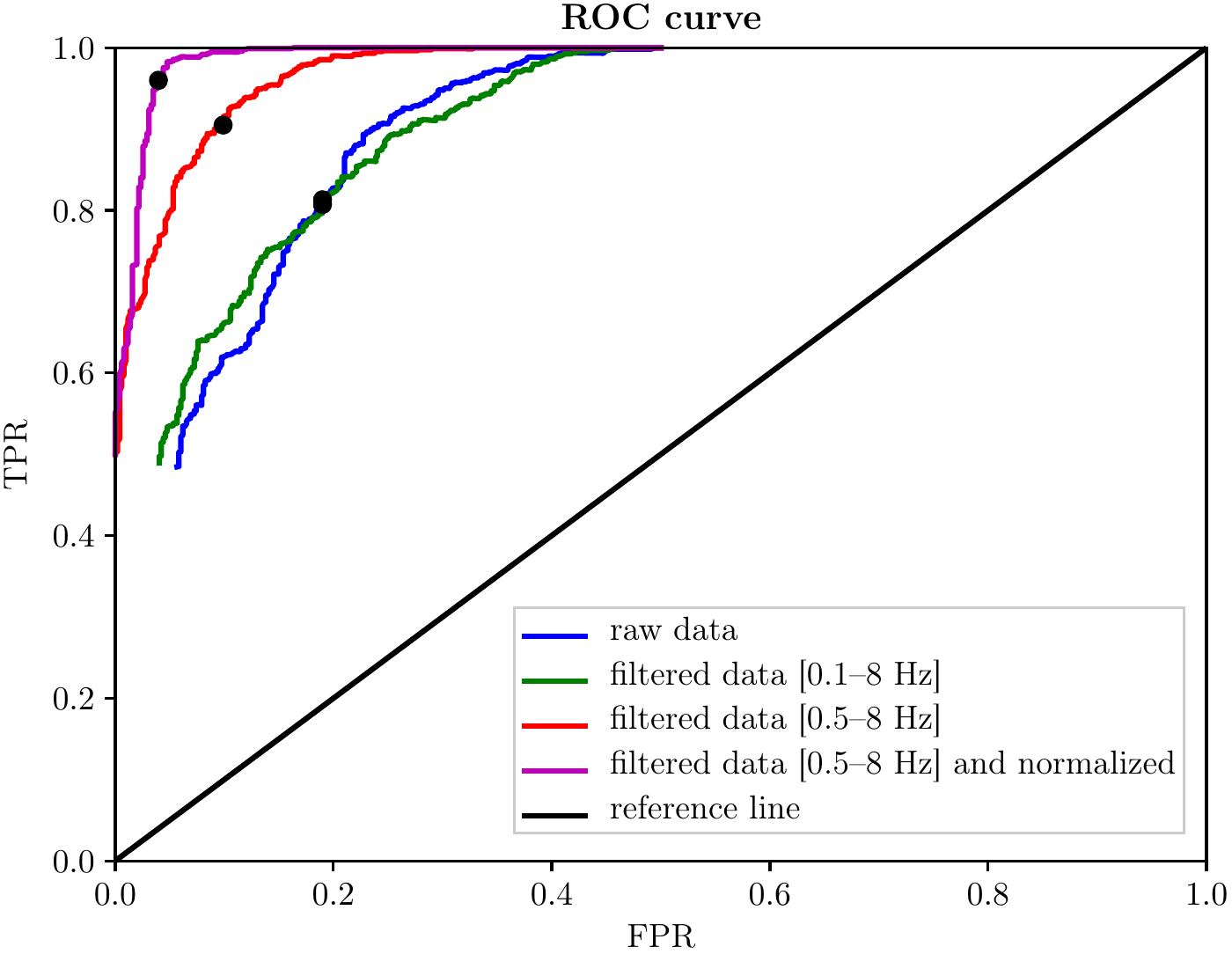}
\label{fig:ROC_EXP2}}
\subfloat[]{\includegraphics[width=1.96in]{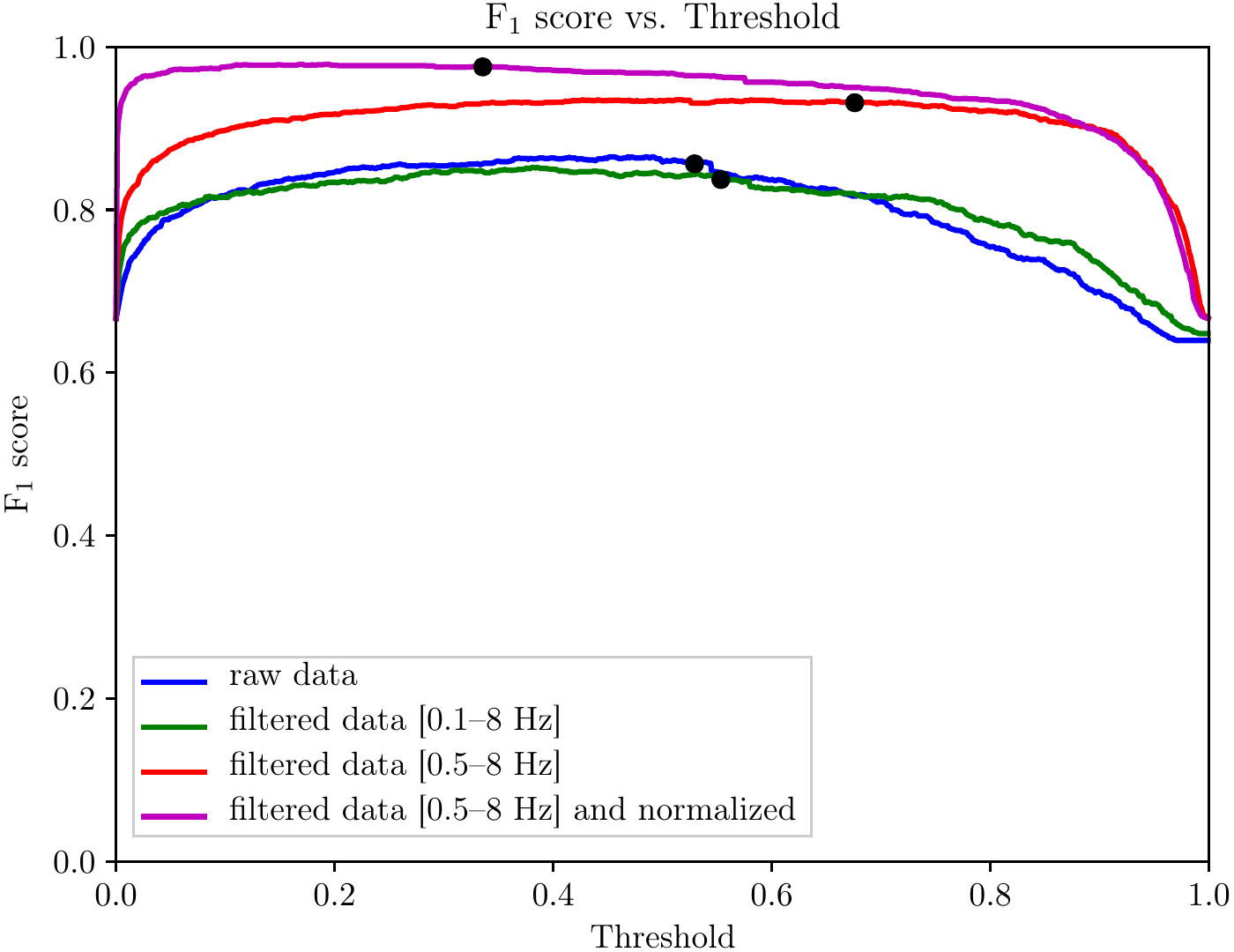}
\label{fig:F1_EXP2}}
\caption{Functional efficiency curves in case leaving 40\% of data out of training. The working points of the EER curve (see  \Cref{fig:EER_EXP2}) are tagged with the symbol {\large\textcolor{black}{\textbullet}}: (a) Precision-Recall curve; (b) ROC curve; (c) F$_1$ score-Threshold curve.}
\label{fig:FCEXP2}
\end{figure}

From the EER curve can be measured the working points for each of the preprocessing modes. These working points can use to obtain other performance measures, as shown in \Crefrange{fig:PR_EXP2}{fig:F1_EXP2}. It can be seen that for raw data and filtered data in the range of 0.1 to 8 Hz, the functional efficiency curves, that is, Precision-Recall, ROC, and F$_1$ score-Threshold curves, behave almost equally for classification purposes. Otherwise, when filtering in the range of 0.5 to 8 Hz is applied to the input data, a qualitative leap is obtained in terms of operational performance, notably about the smooth stability of the F$_1$ score-Threshold curve (see \Cref{fig:F1_EXP2}). Additionally, normalizing the data to $[0,1]$ interval, once filtering the data in the range of 0.5 to 8 Hz, enables the network to operate as a quasi-optimal behavior similar to a perfect classifier. 

\setcellgapes{5pt}\makegapedcells
\begin{table}[ht!]
\caption{Performance metrics for all the input PPG signals modalities use in case leaving 40\% of data out of training. The thresholds refer to the optimal classification thresholds where EER is minimal for each modality (preprocessing) considered.}\label{tab:TEXP2}
\centering
\begin{adjustbox}{max width=\linewidth}
\begin{tabular}{|L|L|L|L|L|}
\hline
\multicolumn{5}{|c|}{\textsc{raw data}} \\
\hline
\multicolumn{1}{|c|}{\textbf{Precision}} &
\multicolumn{1}{|c|}{\textbf{Recall}} &
\multicolumn{1}{c|}{\textbf{F$_1$ score}} & \multicolumn{1}{c|}{\textbf{Threshold}} & \multicolumn{1}{c|}{\textbf{Equal Error Rate (EER)}} \\	
\hline
0.86 & 0.86 & 0.86 & 0.53 & 0.21 \\
\hline
\multicolumn{5}{|c|}{\textsc{filtered data [0.1--8 Hz]}} \\
\hline
\multicolumn{1}{|c|}{\textbf{Precision}} &
\multicolumn{1}{|c|}{\textbf{Recall}} &
\multicolumn{1}{c|}{\textbf{F$_1$ score}} & \multicolumn{1}{c|}{\textbf{Threshold}} & \multicolumn{1}{c|}{\textbf{Equal Error Rate (EER)}} \\	
\hline
0.82 & 0.82 & 0.82 & 0.57 & 0.22 \\
\hline
\multicolumn{5}{|c|}{\textsc{filtered data [0.5--8 Hz]}} \\
\hline
\multicolumn{1}{|c|}{\textbf{Precision}} &
\multicolumn{1}{|c|}{\textbf{Recall}} &
\multicolumn{1}{c|}{\textbf{F$_1$ score}} & \multicolumn{1}{c|}{\textbf{Threshold}} & \multicolumn{1}{c|}{\textbf{Equal Error Rate (EER)}} \\	
\hline
0.93 & 0.93 & 0.93 & 0.68 & 0.11 \\
\hline
\multicolumn{5}{|c|}{\textsc{filtered data [0.5--8 Hz] and normalized in $[0,1]$ interval}} \\
\hline
\multicolumn{1}{|c|}{\textbf{Precision}} &
\multicolumn{1}{|c|}{\textbf{Recall}} &
\multicolumn{1}{c|}{\textbf{F$_1$ score}} & \multicolumn{1}{c|}{\textbf{Threshold}} & \multicolumn{1}{c|}{\textbf{Equal Error Rate (EER)}} \\	
\hline
0.97 & 0.97 & 0.97 & 0.34 & 0.06 \\
\hline
\end{tabular}
\end{adjustbox}
\end{table}

\Cref{tab:TEXP2} shows the experiment's performance metrics, whereby 40\% of data are left out of training. Finally, we compare in \Cref{tab:TCOMP}  performance metrics with other PPG-based biometric methods to consolidate the potential viability attributable to our biometric authentication system. Ratings shown in \Cref{tab:TCOMP} are merely indicative and are limited to the achievements obtained in different experimental scenarios and with different databases. Unfortunately, there is no common roadmap available for the different PPG-based methods to provide the results obtained. However, always with the utmost respect for the work carried out by authors, we chose to report the best performances when there is not enough information available to conduct a comparison that is as fair as possible on equal terms. 

As Spachos \textit{et al.} noted \cite{Spachos2011}, the performance of PPG signal acquisition equipment and the environmental conditions when acquiring the signals impact any biometric authentication system's operational feasibility. So far, most PPG-based biometric systems, as listed in \Cref{tab:TCOMP}, extract the representative features of an individual from the morphology of the PPG signal, either directly from the acquired PPG signal itself or with time or frequency domain transformations. Accordingly, the vulnerability of the morphology of the PPG signal to the physical state of the subject and the environmental and instrumental conditions in the signal acquisition process, restrict its field of application to biometric environments where very stable conditions are guaranteed, namely, when PPG signals, in enrollment and testing phases, were collected under controlled environment and with accurate sensors. 

In the light of the above, the inherent biometric limitations of PPG signal morphology not reflected in the methods collected in \Cref{tab:TCOMP}, where an in-depth analysis reveals the high variability experienced by the parameter EER, degrading its expectations, a priori of the most promising, when PPG signals acquired under different conditions. So, in Sancho \textit{et al.} \cite{Sancho2018}, the range of percentage variation of EER is 13.9 (from 6.9 to 20.8\%) when evaluated on different time-lapses. In Yadav \textit{et al.} \cite{Yadav2018}, the mean EER is 2.82, evaluated on different states and datasets, or in Spachos \textit{et al.} \cite{Spachos2011}, it is 12.75, evaluated on different datasets. In the other methods, only the method's potential is evaluated focusing on the research approach, rather than as a feasible real biometric solution, such as in Karimian \textit{et al.} \cite{Karimian2017}, where the proposed solution provides an Error Rate and rank-1 accuracy of 3.91\% y 99.44\%, respectively, but 8 minutes of PPG signal are required, against the 12 seconds of our approach. Either way, and because all of them use PPG individual cycles, exogenous and endogenous factors in the PPG signal's morphological fluctuations may discourage its use in wearable biometric systems, as consistent and reliable results with proper operations could not guarantee. Our approach holds the best EER of all methods, with a 41\% margin over the second-best result \cite{Yadav2018}. Our method is second in precision, being the best in the comparison \cite{Karimian2017}. Finally, in terms of acquisition and processing time, from all the available time values reported by the studies, our method holds first place with 12.01 seconds. In this sense, it is worth highlighting that our approach does not require new training every time a new user registers; only the user's template pattern to register is needed, which only takes 12 seconds to record.

\setcellgapes{5pt}\makegapedcells
\begin{table}[tbhp]
\caption{The performance of recognition systems based on PPG with state-of-the-art methods compare. Claimed error rates (EERs) involve that in the trial, three attempts were allowed. Acquisition and processing time refers to the system's time to identify whether the user is valid or not.}\label{tab:TCOMP}
	\centering
	\begin{adjustbox}{max width=\linewidth}
		\begin{tabular}{|l|c|c|c|}
			\hline
			\multicolumn{1}{|c|}{\textbf{PPG-based biometric recognition method}} & \textbf{Equal Error Rate (EER) (\%)}  & \textbf{Rank-1 accuracy (\%)} & \textbf{Acquisition and processing time (s)} \\ 
			\hline
			Sancho \textit{et al.} 2018 \cite{Sancho2018} & 6.9 & --- & 21.35 \\ 
			\hline
			Patil \textit{et al.} 2018 \cite{Patil2018} & 23.34 & 86.67 & --- \\ 
			\hline
			Yadav \textit{et al.} 2018 \cite{Yadav2018} & 2.82 & --- & --- \\ 
			\hline
			Karimian \textit{et al.} 2017 \cite{Karimian2017} & 3.91 & \textbf{99.44} & ---\\
			\hline
			Sarkar \textit{et al.} 2016 \cite{Sarkar2016} & --- & 90.53 & 14.00\\
			\hline
			Lee and Kim 2015 \cite{Lee2015} & 3.7 & 96.04 & ---\\ 
			\hline
			Kavsao\u{g}lu \textit{et al.} 2014 \cite{Kavsaoglu2014} & ---  & 94.44  & 13.50 \\
			\hline
			Spachos \textit{et al.} 2011 \cite{Spachos2011} & 12.75 & --- & ---\\
			\hline
			Our approach & \textbf{2.02} & 97.00 & \textbf{12.01}\\
			\hline  
		\end{tabular}
	\end{adjustbox}
\end{table}

The present proposal opens up a new line of work in PPG-based biometry. The study of its diffusion dynamics replaces the analysis of the PPG signal's morphology, our $(p,q)$-planes, highly dependent on the vascular bed's biostructure, an intricate network of tiny blood vessels that branches through body tissues. While deteriorating with age and/or with certain cardiovascular diseases, this vascular microstructure is unique to each individual and maintains a reasonably regular and stable diffusive conductivity over time, making this an excellent biometric marker. Preliminary trials with our biometric authentication system yielded similar performance ratings, with EER and rank-1 accuracy, with one attempt, in the range of about 6\% and 97\%, respectively, when users, initially registered in a  relaxed state, were successfully identified about 30 days later under stress-induced conditions.

\section{Conclusions}\label{sec:CONCLUSIONS}

Over the past ten years, the easily accessible PPG signal has attracted those involved in biometric security. Most PPG-based biometric solutions define the biometric signature out of certain features of the PPG signal morphology. Nevertheless, the high variability of the PPG signal morphology, in reaction to changes in measurement conditions and the individual's psychophysical state, is hampering its adoption as a biometric solution in wearable systems.  

In this research work, still in progress, we propose a robust PPG-based biometric authentication system based on the diffusive dynamics of the PPG signal, arguably very stable in changing environments, instead of morphological aspects of the signal. Our biometrics approach based upon siamese convolutional neural networks, easily integrated into embedded environments that can reach high speeds in the identification process. An error rate, rank-1 accuracy, and enrollment time of 2\%, 97\%, and 12 s, respectively, makes our proposal the best among the eight compared state-of-the-art methods in terms of EER and processing time and the second-best proposal in terms of rank-1 accuracy, indicating a great significance and a potential viability as a real-world biometric system.

With an enrollment time of 12 s, we truly believe that our technical approach can become a real low-cost technological solution. Built-in in miniaturized Tensor Processing Units (TPUs) customized for particular use in wearable biometric systems since once the network has been suitably trained, the authentication methodology does not require successive retraining for reliable serving. Moreover, the memory requirements for storing users' biometric templates, which are around 120 kB, pose no apparent constraints on the authorized user database's portable logistics. With different hardware and software solutions, our efforts aim at reducing PPG signal acquisition time, more in step with the average comparison time, about 10 ms, verifying a user's biometric credentials requesting access to the system. 

Future work involves expanding the dataset with different physiological conditions, but preliminary results with the same individuals under stress conditions and on different days, suggest a promising operational consistency in the authentication process.

\section*{Acknowledgments}
The authors would like to thank Life Supporting Technologies Group (LST-UPM) for taking part in project FIS-PI12/00514, from MINECO.

\bibliographystyle{unsrt}  
\bibliography{Paper4SIAMarXiv}  






\end{document}